\documentclass[referee]{mn2e}
\usepackage{graphicx}

\title[Dynamical friction on Globular Clusters in core-triaxial galaxies.]
{Dynamical friction on Globular Clusters in core-triaxial galaxies: is it
a cause of massive black hole accretion ?}
\author[R. Capuzzo-Dolcetta, A. Vicari]{R. Capuzzo-Dolcetta and A. Vicari\thanks
{E-mail: roberto.capuzzodolcetta@uniroma1.it (RCD); alessandrovicari@uniroma1.it (AV)}\\
Dep. of Physics, University of Roma La Sapienza, P.le A. Moro 5, I-00165, Roma - Italy\\}
\begin{document}

\date{Accepted . Received ; in original form }

\pagerange{\pageref{firstpage}--\pageref{lastpage}} \pubyear{}

\maketitle

\label{firstpage}

\begin{abstract}
The present work extends
and deepens previous examinations of the evolution of
globular cluster orbits in elliptical galaxies, by means of
numerical integrations of a wide set of orbits
in 5 self-consistent triaxial galactic models characterized by a
central core and different axial ratios.
These models constitute a valid and complete representation of
regular orbits in elliptical galaxies.
Dynamical friction is definitely shown to be an efficient cause
of evolution for the globular cluster
systems in elliptical galaxies of any mass and axial ratios.
Moreover, our statistically significant sample of computed orbits
confirm that the globular cluster orbital
decay times are, in a wide range of realistic initial conditions,
much shorter than the age of the galaxies. Consequently, the huge
quantity of mass carried into the innermost galactic region in form
of decayed globular clusters have likely contributed significantly
in feeding and accreting a compact object therein.
\end{abstract}


\section{Introduction}
The effects of dynamical friction (df) of field stars
on the Globular Cluster (GC) orbits have been studied
by many authors, since the pioneering work of Tremaine, Ostriker \& Spitzer (1975).
They deduced that df could have been responsible for
the decay of a significant number of massive globulars into the inner region of
M 31, with possible relevant consequences on the activity of a massive object
therein. In more recent times, Ostriker, Binney \&
Saha (1989) and Pesce, Capuzzo-Dolcetta \&
Vietri (1992) (hereafter PCV) pointed out the role of
triaxiality of the host galaxy in the enhancement of
the effect of dynamical braking.

Actually, in triaxial galaxies the lack of
symmetries in the potential implies the
lack of conservation of any of the components of the
angular momentum, thus df is maximized in
its importance. Indeed, in triaxial
potentials the family of the so called $box$ $orbits$ is present; they
have not a fixed pericenter distance and, so,
objects moving on such orbits can penetrate deeply the inner
regions of the galaxy (they densely fill their permitted region
of the phase space). Box orbits
are not rare, on the contrary they are substantial
in the orbital structure of a triaxial galaxy.
Indeed, it has been shown
(Schwarzschild 1979; Gerhard \& Binney 1985)
that in triaxial potentials, at least those of moderate
axial ratios, the family of box orbits constitutes the $bone$ of
the galactic orbital structure. If we,
additionally, assume the framework of globular
clusters as self-gravitating structures formed during the initial
evolutionary phases of their mother galaxy,
it turns out logical to expect an
initial distribution of GC orbits biased
towards low angular momenta, corresponding to small
pericenter distances (in symmetric potentials) or to
box orbits (in triaxial potentials, see Ostriker et al 1989).

\noindent All these points suggest
as extremely interesting a deep study of the evolution of GC orbits
in axisymmetric potentials as well as in triaxial potentials.
PCV were the first to analyze quantitatively the
role of df in triaxial galaxies, hypothesized to be important first
by Ostriker (1988), on the basis of their simplified scheme.
PCV computed a set of orbits
of GCs in the self-consistent triaxial model developed by
Schwarzschild (1979) and de Zeeuw \& Merritt (1983),
well apt to represent a galaxy of moderate axis ratios
(2:1.25:1). On the basis of the PCV
results, Capuzzo-Dolcetta (1993) enlarged the
data set of orbits in the same potential
and studied the competitive effects of df and of the tidal disruption caused
by a massive galactic nucleus in determining both
the evolution of the radial distribution of GCs and the
quantity of mass lost to the
galactic center in form of orbitally decayed clusters.
The model is validated by that, taking into account both df and tidal effects acting on GCs,
it is able to explain the observed central flattening of the GC radial distribution
in galaxies (see Capuzzo-Dolcetta \& Tesseri 1997; Capuzzo-Dolcetta \& Vignola 1997).
Moreover, the Capuzzo-Dolcetta (1993) results clearly indicate that the quantity of matter carried to the
inner galactic regions in form of massive ($\geq 10^6$ M$_{\sun}$) globulars is relevant,
with relevant consequences on the activity of a massive object in the galactic nucleus.
The actual modes of mass accretion of the massive nucleus by mean of frictionally decayed GCs has been
the object of preliminary investigations (Capuzzo-Dolcetta 2001, 2002, 2003) whose indication is that the supercluster
formed by mean of GCs merged in the inner galactic region may provide enough fuel to
a centrally located black hole to accrete it to a super massive size in a time scale compatible with an
AGN phase.
\noindent
Given this scenario, the aim of this paper is a generalization
to a wide set of triaxial models of different
axial ratios of the results obtained in the case of
just the Schwarzschild's model.

The scheme of the paper is: in Sect. \ref{Model} we
outline the frame of our work, in Sect. \ref{Simulation}
we present the results of our numerical
integrations of both planar and non-planar
orbits of globulars in the various galactic models used;
in Sect. \ref{application} we apply our results
to the study of the evolution of GC systems.
Finally in Sect. \ref{fine} we present a general discussion
and draw some conclusions.

\section[]{The model}
\label{Model}
We study the orbital behaviour
of globular clusters, seen as point-like mass objects
in motion in the regular the regular gravitational force of the
background galaxy and influenced by the df caused by the stars
of the galactic field. Thus, the equation of motion of the GC is:

\begin{equation}
\mathbf{\ddot{r}}_{GC}=-\mathbf{\nabla} V + \mathbf{a}_{df,GC}
\label{eqdiffbase}
\end{equation}

\noindent where ${\bf r}_{GC}$ is the position vector, ${\bf \nabla} V$ is the gradient of
the galactic gravitational potential and ${\bf{a}}_{df,GC}$ accounts for
the frictional deceleration that will be discussed in the next two subsections.

As usual, the second-order vector differential equation of motion is
transformed into a system of first-order differential
equations, to be numerically integrated.
The numerical integration has been performed by mean
of the Runge-Kutta-Merson method (Lance 1960); in particular, we use the
subroutine $dDEQMR$ of the CERN Program Library.
In absence of df, the energy is conserved better than at the 1 per cent level up to 1000 orbital
crossing times. This level of accuracy in the numerical integration is likely conserved when df is
included, as we checked by mean of an independent evaluation of the energy loss (that
obtained solving the differential equation $\dot E={\bf a}_{df,GC}\cdot {\bf v}$, 
where $E$ is the orbital energy per unit mass).

\subsection{Dynamical Friction}
\label{df}
The concept of dynamical friction has been introduced and developed by Chandrasekhar (1943).
The classic Chandrasekhar's formula for the df deceleration term has been 
extended by PCV to the triaxial case, in partial analogy with the Binney's 
extension to the axysimmetric case (Binney 1977), obtaining:

\medskip
\begin{equation}
\mathbf{a}_{df,GC}=-\gamma _{1} V_{1,GC} \mathbf{\widehat{e}}_1-\gamma _{2} V_{2,GC} \mathbf{\widehat{e}}_2-\gamma _{3} V_{3,GC} \mathbf{\widehat{e}}_3
\label{a_df}
\end{equation}
\medskip

\noindent
where $\mathbf{\widehat{e}}_i$ ($i=1,2,3$) are
the eigenvectors of the velocity dispersion tensor and $V_{i,GC}$ is the
component of the GC velocity along the $\mathbf{\widehat{e}}_i$ axis.
The coefficients $\gamma_i$ are (see PCV):

\medskip
\begin{equation}
\label{gammai}
\gamma_i=\frac{2\sqrt{2}\rho (\mathbf{r})G^{2}\ln \Lambda (m+M_{GC})}{\sigma _{3}^{3}} \int_{0}^{\infty} \frac{e^{\Sigma_{k=1}^{3} -\frac{V_{k,GC}/2\sigma _{k}^{2}}{\epsilon _{k}^{2}+u}}}{(\epsilon _{i}^{2}+u)\sqrt{\Sigma_k (\epsilon _{k}^{2}+u)}}du
\end{equation}
\medskip

\noindent
where: $\rho(\bf r)$ is the mass density of
background stars of individual mass $m$, ln$\Lambda$ is
the usual Coulomb's logarithm, $M_{GC}$ is the globular cluster mass, $G$ is the gravitational 
constant, $\sigma_i$ is the eigenvalue, corresponding to $\mathbf{\widehat{e}}_i$, of the velocity dispersion
tensor, and $\epsilon_i$ is the ratio between $\sigma_i$ and the greatest eigenvalue set as $\sigma_3$.

\subsection{The galactic models}
\label{gm}
It is nowadays quite accepted that most elliptical galaxies
and bulges of spirals are at least moderately triaxial in shape
(Franx, Illingworth \& de Zeeuw 1991; Tremblay \& Merritt 1995; Ryden 1996; Bak
\& Statler 2000).
With regard to the shape of the mass density profile, while once was commonly
believed they get flat toward the center, there is now a growing evidence that
the brightness profile of galaxies continue rising.
Actually, the pioneering work of Schweizer (1979) found confirmation in the
HST observations that show how the density distribution, when deprojected,
shows usually a cuspy profile ($\rho \propto r^{-\gamma}$),
at least within the resolution limit of the
instrument (Lauer et al. 1995; Byun et al. 1996; Gebhardt et al. 1996; Faber et
al. 1997, Stiavelli et al. 2001). By the way, available data
show the existence of two well defined classes of density profiles.
Brighter, more massive, ellipticals show a brake in the luminosity vs radius
profile, that gets very shallow (slopes in the range $0< \gamma < 0.3$)
at the few parsec scale (``core" galaxies in the sense of Lauer et al. 1996);
fainter ellipticals have steep power-law profiles ($\gamma \ga 1$) that continue
essentially unchanged in to the resolution limit.

For the purposes of this work, it is necessary to have
both the matter density profile and the velocity distribution
(i.e. the velocity dispersion tensor in Eq. \ref{gammai});
unfortunately, these latter data are not yet available
for the models of self-consistent triaxial galaxies with density cusps
developed so far (Merritt \& Fridman 1996; Holley-Bockelmann et al. 2001).
For this, we decided to rely on the large set of self-consistent triaxial
models presently available (Statler 1987 and Statler private communication),
that are based on the so-called ``perfect ellipsoid"
models, i.e. those given by the mass density:

\begin{equation}
\label{rho_r}
\rho(m) =\frac{M}{\pi^2abc} \frac{1}{\left[ 1+m^2 \right] ^{2}},
\end{equation}
where
\begin{equation}
\label{m}
m^2=\frac{x^{2}}{a^{2}}+\frac{y^{2}}{b^{2}}+\frac{z^{2}}{c^{2}}.
\end{equation}

\noindent
($a> b> c$) and M is the galactic mass.
This family of density profiles has a central core,
whose spatial size is $\approx 0.64a$, so that, in principle,
they should be considered a good modelization of real massive core
galaxies only.
At the same time these profiles, at large galactocentric distances,
have the same $m^{-4}$ behaviour of the cuspy Merritt \& Fridman (1996)
and Holley-Bockelmann et al. (2001) models
(triaxial generalizations of the Hernquist 1990 and the Dehnen 1993 models),
that do not account for the extended dark matter component that should be more
gently decreasing.
As a matter of fact, with regard to the study of dynamical friction,
these approximations in the representation of the galactic profile in the
innermost and outermost regions correspond to an underestimate of its effect.
Thus we feel confident that the results of this paper represent, at least,
valid ``lower limits" for all the dynamical friction-induced effects also for
cuspy galaxies with a dark matter halo.

Moreover, the potential generated by the perfect ellipsoid
has many advantages, as we now describe.
Actually, the solution of the Poisson's equation for the
density law (\ref{rho_r}) is a potential in the so called
St\"{a}ckel form (Weinacht 1924; de Zeeuw 1985).

\begin{eqnarray}
V(x,y,z)=-G \frac {M} {\pi} \int_0^{\infty} \frac {du} {(1+ \frac {x^2}{a^2+u}+ \frac{y^2}{b^2+u}+\frac{z^2}{c^2+u}) \sqrt{(a^2+u)(b^2+u)(c^2+u)}}
\label{potenziale}
\end{eqnarray}

\vspace{0.4cm}
\noindent that is fully integrable, because the
Hamilton-Jacobi equation separates with the choice of
the cofocal ellipsoidal coordinates
(denoted with $\lambda$, $\mu$, $\nu$).
The model represented by the density law (\ref{rho_r})
is called ``perfect ellipsoid" because it is the unique mass
model having the potential in a St\"{a}ckel form
in which the density is stratified on concentric
ellipsoids (de Zeeuw \& Lynden-Bell 1985).
In addition to being fully integrable, the potential (\ref{potenziale})
is characterized by that all the orbits are regular, i.e. they respect
three integrals of motion: the usual energy integral ($E$) and two
non classical integrals ($I_2$, $I_3$) that converge, in the spherical
and axisymmetric limits, to the conserved components of the angular momentum.
The results obtained by mean of the perfect ellipsoid
model (whose orbital properties have been fully studied
by de Zeeuw 1985) constitute a deep analysis of
the behaviour of regular orbits in triaxial potentials and
give, also, results of astrophysical interest for
what regards elliptical galaxies with a central density core.
Actually, even if in the cuspy models chaotic
(irregular) orbits are in a relevant fraction (Gerhard \& Binney 1985;
Merritt \& Fridman 1996; Merritt \& Valluri 1996), some results
seem to indicate that their presence does not affect significantly
the orbital decay of the satellite (Cora, Vergne \& Muzzio 2001).

\begin{table*}
\begin{center}
\caption{Galactic model characteristics. Models are numbered as in Statler (1987).
$a$, $b$ and $c$ are the axes and $T=(a^2-b^2)/(a^2-c^2)$
is the triaxiality parameter (Statler 1991),
$E_{min}$ is the depth of the potential well and $E_2$ and $y_{_{lim}}$ are the
minima values of the initial energy and y-starting point, respectively, that
generate a loop orbit.
$E_{min}$ and $E_2$ are in units of $[GM^2a^{-1}]$.
}
\label{modelli}
\begin{tabular}{|l|c|c|c|c|c|c|} \hline \hline
Model   &b/a    &c/a    &$T$   &$E_{min}$ &$E_2$    &$y_{_{lim}}/a$\\
 \\ \hline
04      &0.625  &0.500  &0.81  &-0.9074   &-0.6660  &0.78\\ \hline
06      &0.875  &0.500  &0.31  &-0.8098   &-0.7330  &0.48\\ \hline
16      &0.250  &0.125  &0.95  &-1.5296   &-0.6606  &0.96\\ \hline
19      &0.625  &0.125  &0.62  &-1.1347   &-0.7766  &0.78\\ \hline
21      &0.875  &0.125  &0.24  &-0.9855   &-0.8747  &0.48\\ \hline \hline
\end{tabular}
\end{center}
\end{table*}

The self-consistent problem of the ``perfect ellipsoid"
has been numerically solved by Statler (1987), who
found a solution for all the explored range of axis
ratios, suggesting its existence for any axis ratio.
Statler found the numerical solution for 21 different
values of the axis ratios using an unbiased catalog
of 1065 orbits and matching the density of
the mass model at 240 grid points. The grid has been constructed
dividing the space with 15 shells in $\lambda$, and with four
divisions for both $\mu$ and $\nu$.
Statler (private communication) recomputed 5 out of the 21 different
models of the 1987 paper with a finer grid (960 shells,
constructed with 8 divisions both in $\mu$ and in $\nu$)
and using the improved Lucy's method described in Statler (1991).
The main characteristics of these models are reported in Table \ref{modelli}.

\section{Simulation of the orbital decay}
The role of df in the core-triaxial galaxies modeled
as described in Sect. \ref{Model} is studied here by numerical
integration of a wide set of GC orbits.

Hereafter, we assume that all the
quantities are expressed taking as length and mass units $a$ and $M$,
respectively. Moreover, setting $G=1$ implies

\begin{eqnarray}
\tau=G^{-1/2} M^{-1/2} a^{3/2}=1.49 \cdot 10^6  \bigg{(}\frac{M}{10^{11}{\rm M_{\sun}}}\bigg{)}^{-1/2}
\bigg{(}\frac{a}{1 \: {\rm kpc}}\bigg{)}^{3/2}  \: yr \:
\label{tauadimen}
\end{eqnarray}

\noindent
as unit of time.

The linear scaling of the df deceleration with $M_{GC}$ (when $m<<M_{GC}$,
see eq. \ref{gammai}) reflects into
an inversely proportionality between the time needed by the satellite
to sink to the centre and $M_{GC}$. We exploit this to reduce the computational time,
because we can fix a quite high mass for the GC, without any loss of generality.
By the same token, the choice ln$\Lambda=10$ (i.e. a value that can be considered
as typical for a massive object in a galactic environment)
is not limiting, because of the linear scaling with ln$\Lambda$ of the df
braking term.
In the following, if not explicity mentioned, we, indeed,
refer always to an orbital evolution of GC with $M_{GC}=5 \times 10^{-4} M$;
this corresponds to
$M_{GC}=5\times 10^7$ M$_{\sun}$ for a typical elliptical galaxy with
$M=10^{11}$ M$_{\sun}$.

\label{Simulation}

\subsection{The choice of the orbits}
\label{Choice}

\begin{figure}
\begin{center}
{\includegraphics[angle=0,scale=0.6]{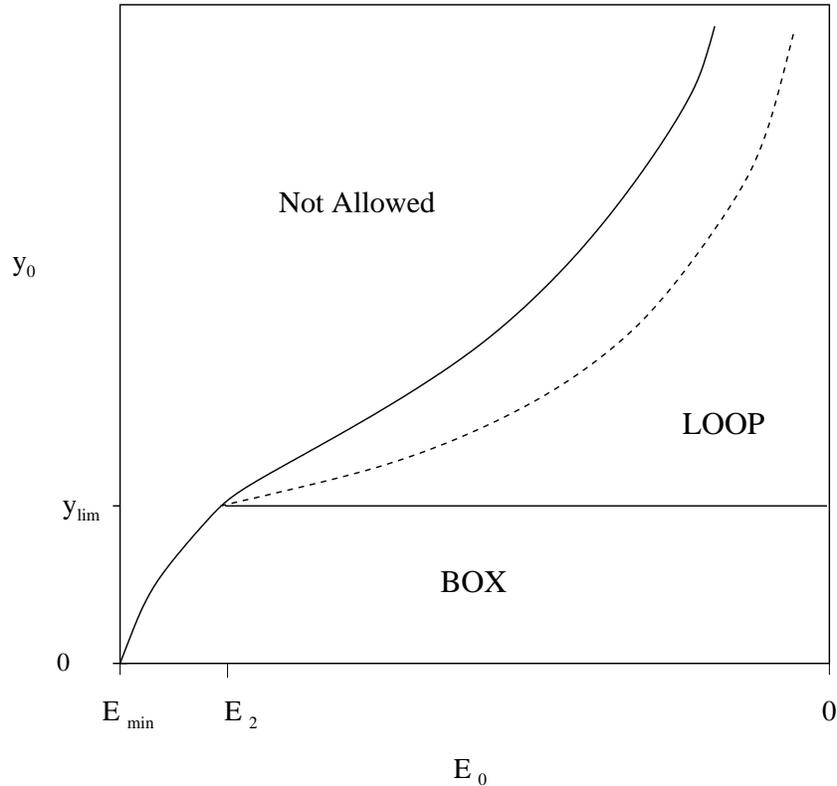}}
\caption{Curves delimiting the regions of the $(E_0,y_0)$ plane where box and
loop orbits are permitted. The dashed curve is $y_{0,cl} (E_0)$, that
corresponds to the initial values of $y_0$
giving rise to the particular case of closed, periodic loops. }
\label{ci}
\end{center}
\end{figure}

We have performed numerical integrations of many ($\sim 850$) bound orbits
($E_{min}\leq E <0$) of different types ($radial$, $box$ and different kind of $loop$
orbits of different elongation)
at various energies, in the ranges $0.05 \leq E/E_{min} \leq 0.95$
for planar box orbits, $0.05 \leq E/E_{min} \leq E_2/E_{min}$ for planar loop orbits
($E_2$ is the limiting energy for planar loop orbits,
see Fig. \ref{ci} and Table \ref{modelli})
and $0.05 \leq E/E_{min} \leq 0.6$ for all non-planar orbits
\footnote{$E/E_{min}=0$ corresponds to the asympotically free orbit,
while $E/E_{min}$=1 to a GC at quiet in the origin.}.
The choice of the lower limit ($0.05$) for $E/E_{min}$
is done to avoid an exceedingly large relative error in
the energy as merely due to a value of energy too close to zero.
On the other hand, the choice of the $0.95$ limiting value is to avoid
a useless integration of orbits already confined in the
very inner regions; the $0.6$ limit for non planar orbits is sufficient to
have an adequate sampling of all the four orbital families.

Simulations were stopped at the ``orbital decay time", $T_{df}$,
defined as the first instant when the following two conditions are fulfilled:
$i)$ the GC energy is reduced to a value such that
$|(E-E_{min})/E_{min}| < 10^{-4}$;
$ii)$ the df energy decrement in a time step is
comparable to the numerical error in the energy.
Condition $i)$ guarantees that the GC is confined in the
innermost galactic region, while the second condition ensures
the motion is eventually limited to an oscillation
whose amplitude decrease cannot be furtherly followed due to
the overwhelming roundoff error.
In all the cases studied these conditions correspond
to apocenter distances less than $0.01a$.

To select orbits, we exploit that all the orbits in a triaxial
potential cross at least once the $y$ (or $x$) axis with the velocity
vector perpendicular to that axis (Schwarzschild 1982).
Thus, for planar orbits, we always start the integration with the GC
at ${\bf r_0}$=$(0,y_0 \geq 0,0)$, and
with ${\bf v_0}$=$(v_{x0}\geq 0,0,0)$ choosing $y_0$ and $v_{x0}$
such to obtain the required initial energy ($E_0$).
With this choice of the initial conditions we can easily classify in 
all the galactic models studied the morphology of the planar orbits 
in the following way:
for any given energy $E_0$, box orbits have $y_0$ smaller than
a fixed value $y_{_{lim}}$ (a parameter that depends on the
axis ratios of the models, see Table \ref{modelli});
just above $y_{_{lim}}$ we find the most elongated loop orbits.
A further increase of $y_0$ results in less elongated orbits up to the
value, $y_{0,cl}$, corresponding to the (unique) closed orbit,
that is quasi-circular at high energies. Above
this value we re-obtain the same (previously found) elongated orbits
($i.e.$ loop orbits cross the $y$ axis with the velocity
vector perpendicular to the axis itself twice). To resume,
Fig. \ref{ci}, that refers to a generic model of Table 1,
represents a paradigma to generate all the possible planar orbits.

\noindent Also to obtain non planar orbits we fix the initial energy
and the initial position on the
$y$ axis ($x_0=0,y_0 \geq 0,z_0=0$): then we choose
the initial velocity orthogonal to the $y$ axis,
pointing in the positive $x$ and $z$ directions
($v_{x0}=v_0cos \phi >0, v_{y0}=0,
v_{z0}=v_0sin \phi>0 $),
letting the inclination angle $\phi$ over the $x-y$ plane
assume the five values $15^\circ$, $30^\circ$, $45^\circ$,
$70^\circ$ and $90^\circ$ (the latter value corresponds to
an orbit in the $y-z$ plane).
\par\noindent This way, a sample of orbits
sufficiently wide to draw conclusions about differences
in the evolution of orbits limited to a plane
and moving in 3D is obtained.

We checked that the particular choice of the initial condition
influences slightly the orbital decay time (less than 5\% for the highest
energies and even less for lower energies),
in the sense that two GCs starting on two different points of a given orbit
have very similar decay times. This ensures that the results on
the role of df presented here are quite general.

\begin{figure}
\begin{center}
{\includegraphics[angle=0,scale=0.5]{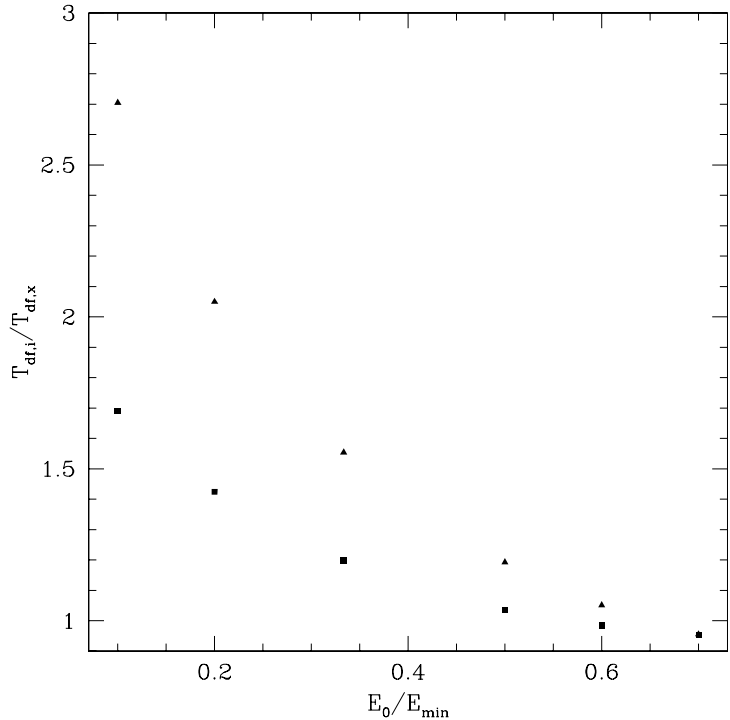}}
\caption{The ratios of the df decay times
of quasi-radial orbits along the $y$ (squares) and $z$
(triangles) axis to that of orbits in the $x$ direction,
as functions of the initial orbital energy.
Results are relative to model 04.}
\label{rad}
\end{center}
\end{figure}

\begin{figure*}
\resizebox{\hsize}{!}
{\includegraphics[angle=0]{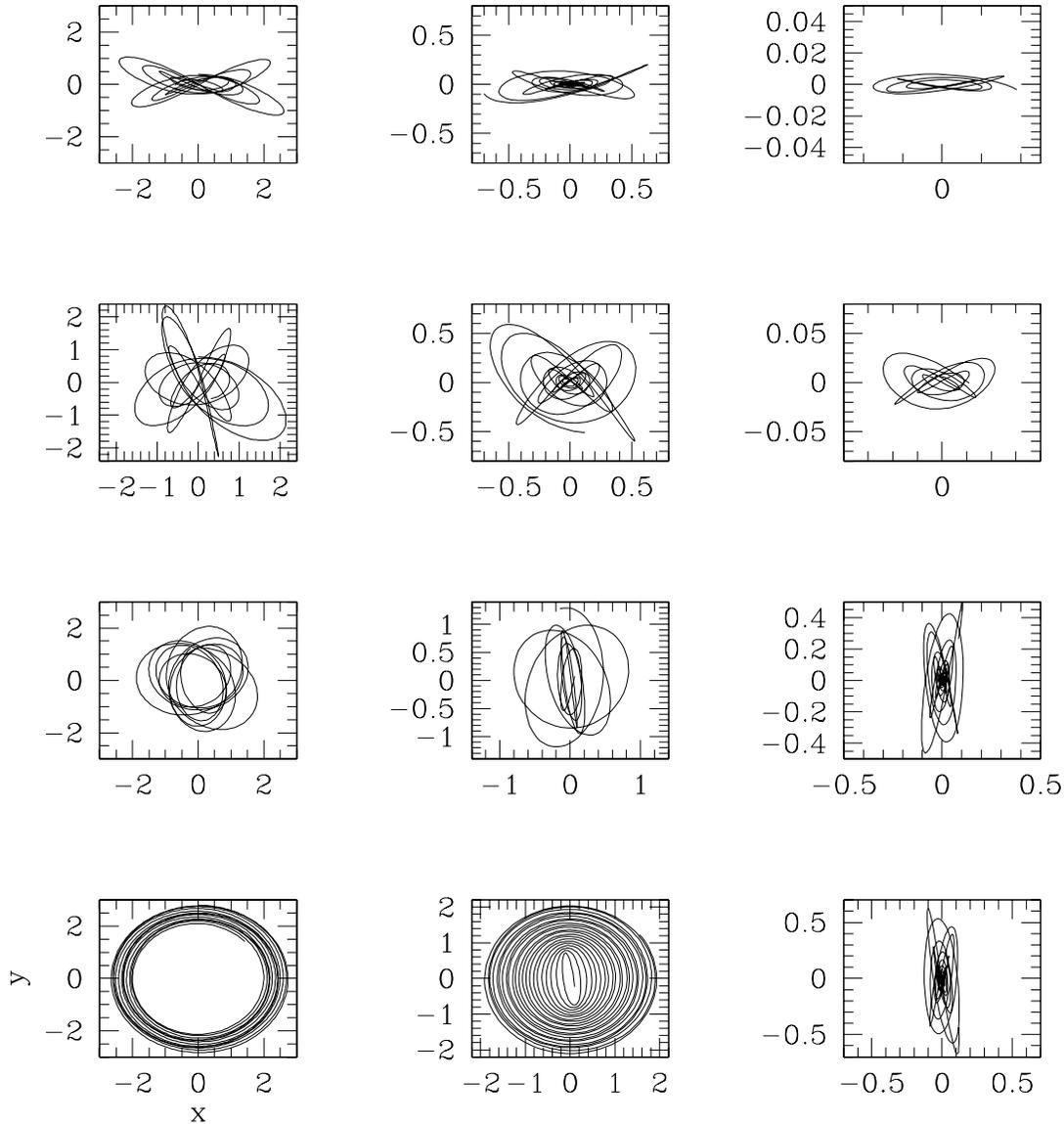}}
\caption{Every column represents four time snapshots
corresponding to an initial, an
intermediate and a final phase of the evolution
(time flows left to right) of four planar orbits subjected to dynamical
friction in the case of model 04.
Top to bottom: a typical (thin) box; one of the least elongated box;
a typical (elongated) loop; a closed loop.}
\label{tevol}
\end{figure*}

\begin{figure}
\resizebox{\hsize}{!}
{\includegraphics[angle=0]{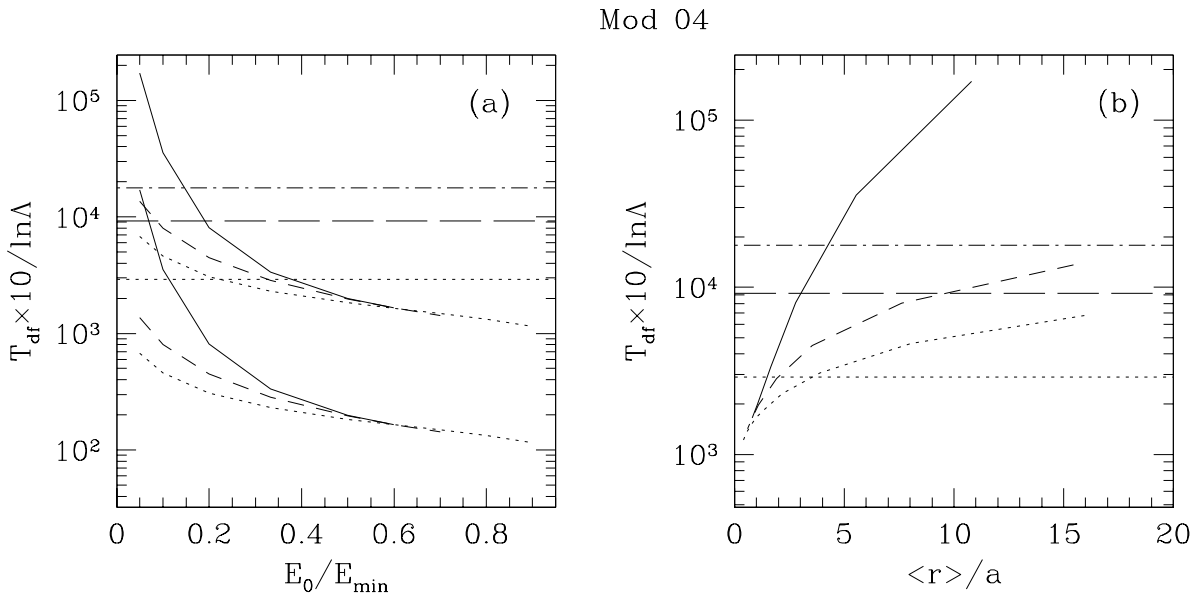}}
\caption{$Panel$ $(a)$: the energy decay times of a GC with $M_{GC}=5 \times 10^{-5} M$
(upper curves) and of a GC with $M_{GC}=5 \times 10^{-4} M$ (lower curves)
as functions of the initial energy.
$Panel$ $(b)$: the energy decay times of a GC
with $M_{GC}=5 \times 10^{-4} M$ vs the mean radius of the corresponding {\it unperturbed}
(i.e. without df) orbit.
Both panels refer to model 04 and different curves refer to
the different kinds of planar orbits (solid:
$closed$ $loops$; dashed: $least$ $elongated$ $boxes$;
dotted: $x-radial$ orbits). Horizontal lines
mark the age of the universe (assumed 13.7 Gyr, as from the recent determination
by Spergel et al. (2003) on the basis of the WMAP satellite data) corresponding to three choices of $a$ and
$M$: dot-dashed is for a typical dwarf elliptical ($a=0.1$ kpc, $M=10^7$ M$_{\sun}$),
dashed for a normal elliptical ($a=1$ kpc, $M=10^{11}$ M$_{\sun}$)
and dotted for a compact bulge ($a=0.2$ kpc, $M=3 \times 10^9$ M$_{\sun}$).
}
\label{plan1}
\end{figure}

\begin{figure}
\resizebox{\hsize}{!}
{\includegraphics[angle=0]{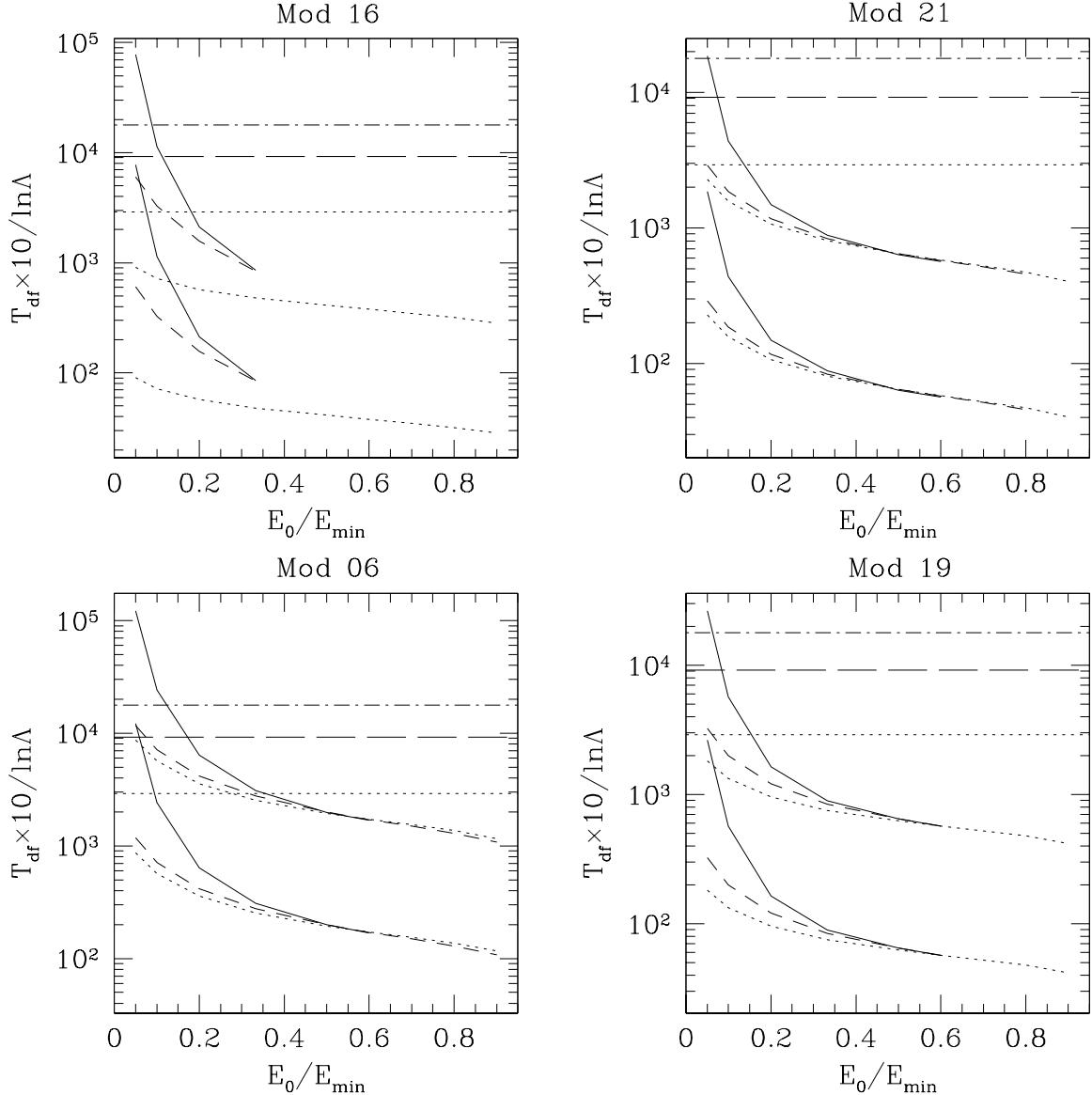}}
\caption{Energy decay times for the models 6, 16, 19 and 21 of
Table \ref{modelli} as functions of the initial energy. All symbols are as in Fig. \ref{plan1}.}
\label{plan2}
\end{figure}

\begin{figure}
\begin{center}
{\includegraphics[angle=0,scale=0.6]{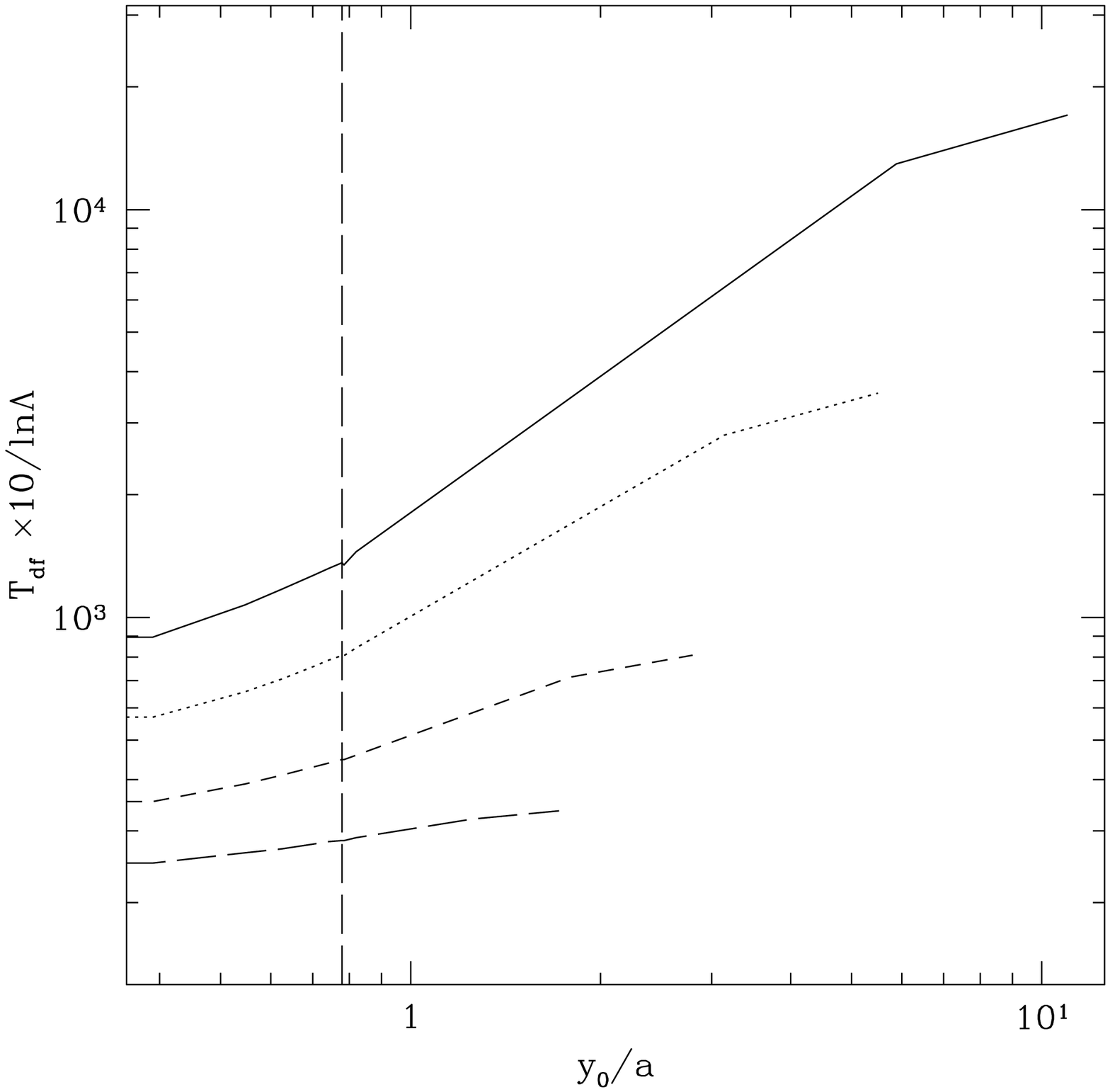}}
\caption{The dependence of $T_{df}$ on the initial condition $y_0$ for four
different $E_0$ in the model 04. $E_0/E_{min}$=0.05, 0.1, 0.2 and 1/3 correspond to the solid,
dotted, short dashed and long dashed line, respectively.
The vertical line indicates the value of $y_{lim}$.
}
\label{y0vstf}
\end{center}
\end{figure}

\subsection{Quasi-radial orbits}
\label{Radial}
To understand better the characteristics of the evolution of orbits in
presence of df it is useful to have a glance at radial orbits.
Obviously, in a triaxial galaxy, radial orbits are a-priori allowed
along the axes, only.
We checked that in our models radial orbits are stable
for any initial energy just along the major axis ($x$);
those along $y$ and $z$ are stable at low energies, only,
while for higher energies they quickly develop an $x-$oscillation.
The decay times of orbits along the 3 axes are similar at low
energies, when the orbits are quasi-radial along all the 3 axes,
while they increasingly differ at higher energies (see Fig. \ref{rad}).
Actually, the efficiency of df as braking
mechanism is greater for greater values of the phase-space density
which roughly follows the behaviour of the space matter density
$\rho({\bf r})$ (see PCV); as a matter of fact, moving
along an axis of simmetry, $\rho({\bf r})$ is flatter (greater) in the direction of greater
lenght scale parameter ($a>b>c$) at a given distance from the origin and
high enough energies are required for a GC to sample a spatial
region where the density difference is appreciable along the various axes.

\subsection{Planar orbits}
\label{Planar}
To resume the general characteristics of the shape-evolution of orbits subjected
to df, we show in Fig. \ref{tevol} some time-snapshots of
four representative orbits: a closed loop, a typical (elongated loop) an almost
``squared" box and an elongated (typical) box.
The evolution of the closed loop orbit (initially quasi-circular) shows an
increasing collimation in the $y-$direction (bottom row of Fig. \ref{tevol}).
This is due to indirect $(i)$ and direct $(ii)$ effects of df:
$(i)$ the initial reduction of the orbital
energy due to df braking trasforms the quasi-circular shape into that of a
typical low-energy loop, which is characterized by being
elongated in the $y-$direction (de Zeeuw, 1985);
$(ii)$ a further reduction of energy leads, eventually, the
orbit to become a very thin (quasi-radial) box. This happens at an
energy greater than $E_2$ (the minimum energy allowed to
loop orbits), thus we consider the latter
a direct effect of df because
it corresponds to an actual change of the type of orbit.
An effect of df different in the two coordinate
directions was also expected simply by that $a>b$,
that corresponds to a matter density distributed
flatterly along the $x$ direction, meaning a stronger
deceleration along $x$.
The evolution of the typical (elongated) loop
orbit shows the same general behaviour of the closed loop (like the progressive
alignment along the y-axis).
Moreover, we note that the evolution of both the closed and elongated loop orbits
consist of a sequence of orbits keeping the same initial general
shape characteristics until they turn into boxes.
Indeed also the elongated loop eventually decays as a low energy box,
slightly less aligned along the $y-$axis than the closed loop.

\noindent The time evolution of
both the box orbits in Fig. \ref{tevol} is mainly that of shrinking in the two
directions, even with a slight flattening along the $x-$direction.
The explanation for this is that df causes an evolution of the orbit toward
low energies and, in the $x-y$ plane, stable low-energy orbits are usually
more extended along the $x$ direction. This effect overcomes that due to the
larger friction along that axis as expected on the basis of the positive
correlation between the density along an axis and the
efficiency of df for (quasi) radial orbits discussed at the end of SubSect. 3.2.

The quantitative results of our simulations are summarized
in Figs. \ref{plan1} and \ref{plan2} that show the
dependence of the decay times of the various families of orbits
upon their initial energy and also (just for the model 04) upon the mean radius of
the corresponding unperturbed (i.e. without df) orbit.
Fig. \ref{plan1} shows that the scaling of the decay time with the mean
galactocentric radius of the corresponding unperturbed orbit is
different for the different families of orbits, being
always an increasing function of $<r>$.

As evident from Figs. \ref{plan1} a) and \ref{plan2}, for any initial energy
closed loop orbits have longest decay times, while the $x-$radial orbits have them 
shortest.
Because our orbital integrations show that box orbits with larger $y_0$ have
longer decay times (as shown by the positive correlation between $T_{df}$ and $y_0$ for
given initial energy $E_0$ of Fig. \ref{y0vstf}), it is possible to conclude that all
box orbits are confined between the dotted and the dashed lines in the figures \ref{plan1}-\ref{plan2}.
Analogously, we found that loop orbits with smaller values of $y_0$
($y_{lim}\leq y_0\leq y_{0,cl}$, see discussion of Fig. 1)
are more powerfully decelerated than closed loops with the same
initial energy. This because the presence of the density term in the
expression of df deceleration makes its effect more evident
in the inner regions of the galaxy where
elongated orbits plung. Consequently,
all loop orbits should be confined between the dashed
and the solid lines in Figs. \ref{plan1}-\ref{plan2}; it is worth noting that there
are, instead, few very eccentric loop orbits that do not
follow this rule and that decay in the innermost galactic region in a
time shorter than that of few less elongated box orbits. Anyway,
we found that just loops with $y_0$ in a very narrow range above $y_{lim}$
decay faster than box orbits. The width of the range is $\leq 0.005y_{lim}$, so that
the number of loop orbits evolving faster than box orbits is almost negligible.

\begin{figure}
\resizebox{\hsize}{!}
{\includegraphics[angle=0]{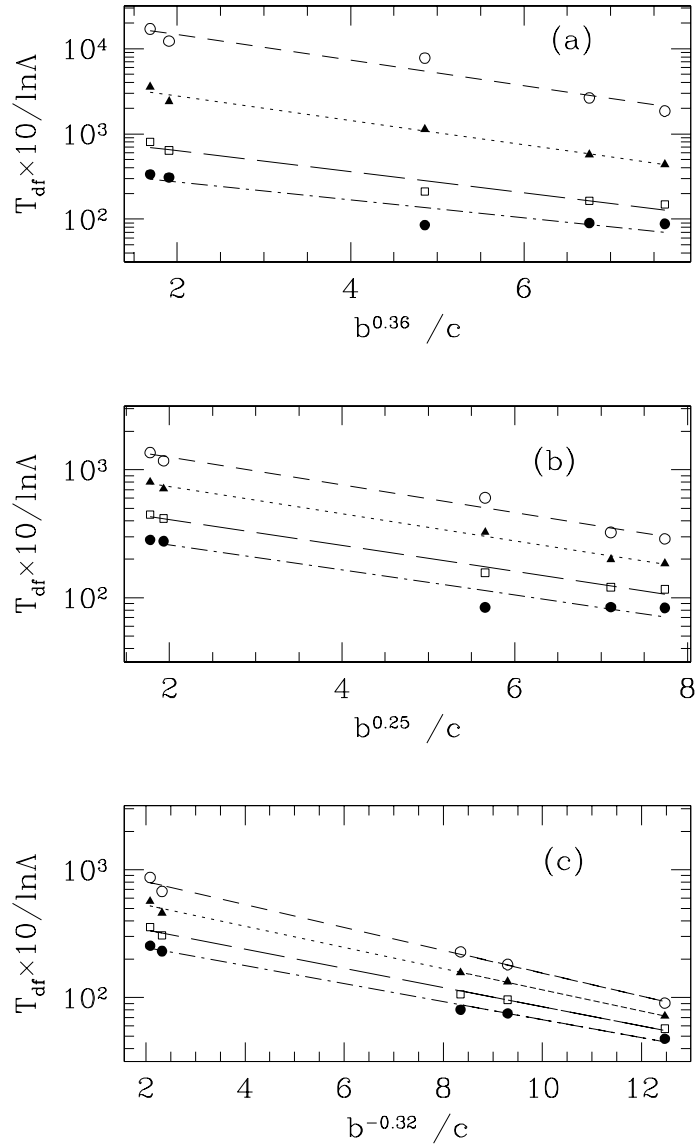}}
\caption{Energy decay times as function of the parameter $b^{n}/c$
for different types of orbits: closed loops (panel $(a)$),
boxes (panel $(b)$) and $x-$radial orbits (panel $(c)$).
In every panel, different symbols identify different initial
orbital energies: $E_0/E_{min}=0.05$ (empty circles), $0.1$
(black triangles), $0.2$ (empty squares), $1/3$ (black circles).
The straight lines are least squares fits to data.
}
\label{bsuc}
\end{figure}

One of the scopes of this paper is the exam of the dependence of
the df decay time on the axial ratios of elliptical galaxies.
In the galactic models used here, for a given
total galactic mass there is an obvious
inverse linear dependence of the df efficiency on the axes lengths
through the central density at fixed galactic mass ($\rho_0=M/(\pi^2abc)$) that is
partly counterbalanced by that increasing $a$, $b$ or $c$ the size
of the core increases and df is larger in larger cores (this dependence
on the mass density is just sligthly modified by
variations of the velocity dispersion tensor with $a,b$ and $c$).
We found a quasi-linear dependence of Log$T_{df}$ on $b^n/c$,
as clearly shown in Fig. \ref{bsuc}, for the whole range of
orbital energies investigated, with $n=0.36$ for closed loop orbits,
$n=0.25$ for the lesser elongated boxes and $n=-0.32$ for $x-$radial orbits.
The negative value of $n$ (that implies larger values of $\rho_0$ for larger
$b^n/c$) for the $x-$radial orbits is due to that
the ``central" density effect overcomes the other effect mentioned above, because
these orbits travel along the $x-$axis where, obviously, an effect
depending on the length scales of the other two coordinate axes is negligible.

\subsection{Non planar orbits}
\label{Nonplanar}
Of course, a reliable application of the
results to the study of the orbital evolution
of GCs in presence of df needs the computation of
a large set of orbits sampling as
completely as possible the allowed phase-space. In
Subsect. \ref{Planar} we have
deeply examined the orbital evolution in one
of the coordinate planes; obviously, this is not an exhaustive representation of
all the possible orbits of clusters in their host galaxy but it
suffices to outline some general conclusions on the
role of df in the global evolution of Globular Cluster Systems (GCSs)
in galaxies of various size, mass and axis ratios.
We postpone to a following paper a thorough investigation of the df effects on the
evolution of a GCS in an elliptical galaxy by mean of the analysis
of a complete sample of orbits, limiting here ourselves to results concerning some
selected 3D orbits.

\begin{figure}
\resizebox{\hsize}{!}
{\includegraphics[angle=0]{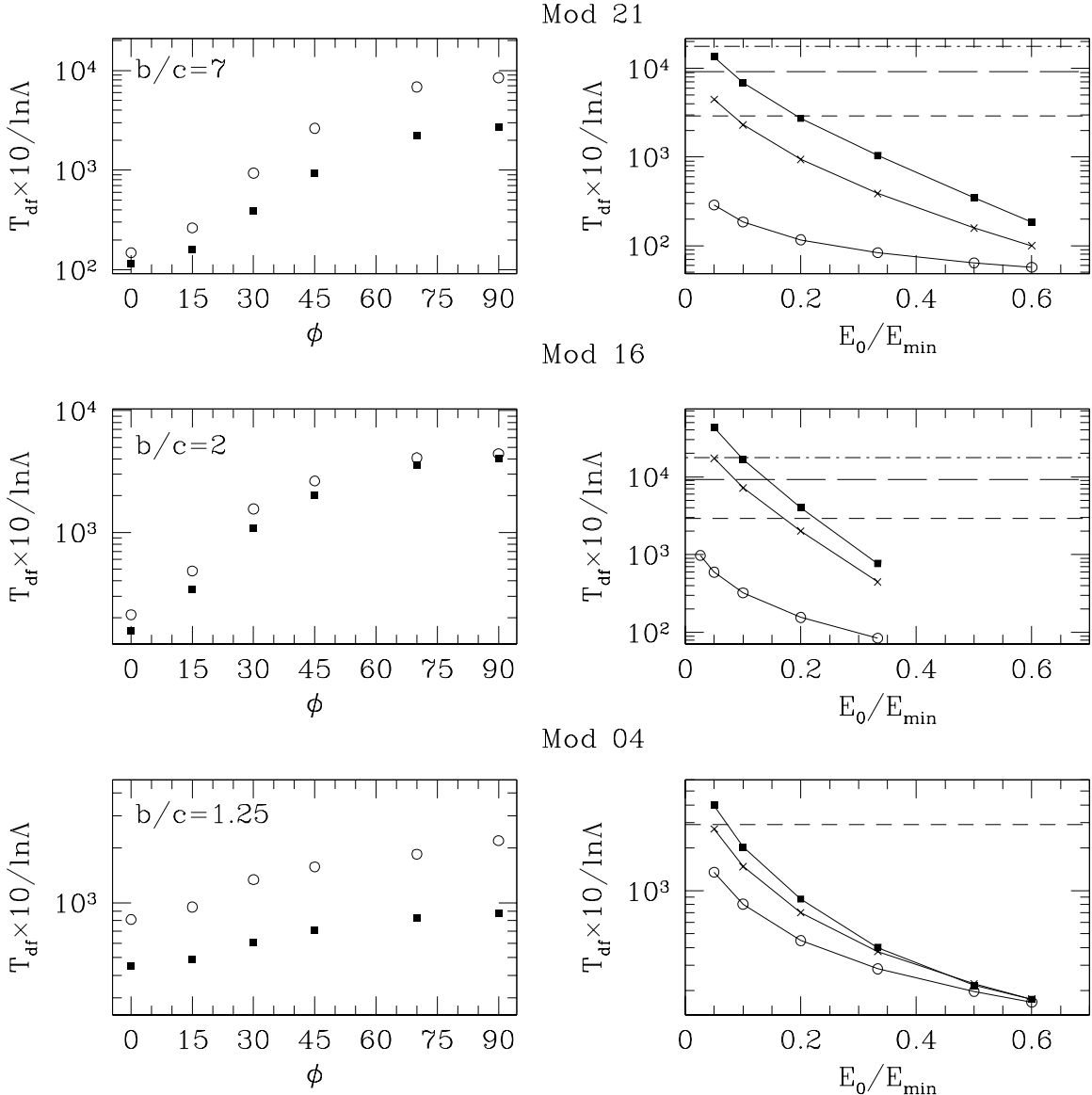}}
\caption{
Left column: energy decay time vs the initial
inclination angle $\phi$ (in degrees) for three galactic models
(models 04, 16 and 21, bottom to up). Black squares represent
orbits starting with $y_0=y_{_{lim}}$ and open circles
those with $y_0$ corresponding to that of
the closed loop ($y_0=y_{_{0cl}}$);
the initial orbital energy $E_0$ is set to 0.2 $E_{min}$.
Right column: the dependence of the energy
decay time upon the initial energy for orbits at 
different angles $\phi$ over the $x-y$ plane.
Orbits have $y_0=y_{_{0cl}}$. For the sake of presentation,
we report just the $\phi =0^\circ$ (empty circles),
$45^\circ$ (crosses) and $90^\circ$ (filled squares) cases.
Horizontal lines in the right panels mark the age of the universe,
as in Fig. \ref {plan1}.
}
\label{3d}
\end{figure}

Fig. \ref{3d} shows the energy-dependence
of the df decay time for 3D orbits.
The decay time is increasing with the initial inclination
angle over the $x-y$ plane, making the $y-z$ planar orbits a
lower limit to the df efficiency;
this is due to that the $z-$axis
has the shortest space scale parameter
($c$). Correspondingly, the density along
this axis decreases more steeply than
along $y$ and $x$ and df is thus maximized on the $x-y$ plane.
A remarkable feature of Fig. \ref{3d} is that the
ratio between the $T_{df}$ value of an orbit moving on the $y-z$
plane ($\phi=90^\circ$) to that of an orbit
of the same energy (and type) limited to the
$x-y$ plane ($\phi=0^\circ$) is much larger
for models with larger $b/c$ ratios
(indeed, for fixed $a$ we already noticed that a larger axis
length implies a shorter $T_{df}$).

\noindent
Because of this, the modelization (that we present
in the next section) of a GCS as
composed just by objects moving on
the $x-y$ plane leads to a slight overestimate
of the role of df on the GCS evolution.
A quantitative evaluation of this overestimate
is straightforwardly obtained by Fig. \ref{3d}
itself because, by the same arguments above,
$y-z$ is the plane where df is minimized.

\section{An application to the evolution of Globular Cluster Systems}
\label{application}

\begin{figure}
\resizebox{\hsize}{!}
{\includegraphics[angle=0]{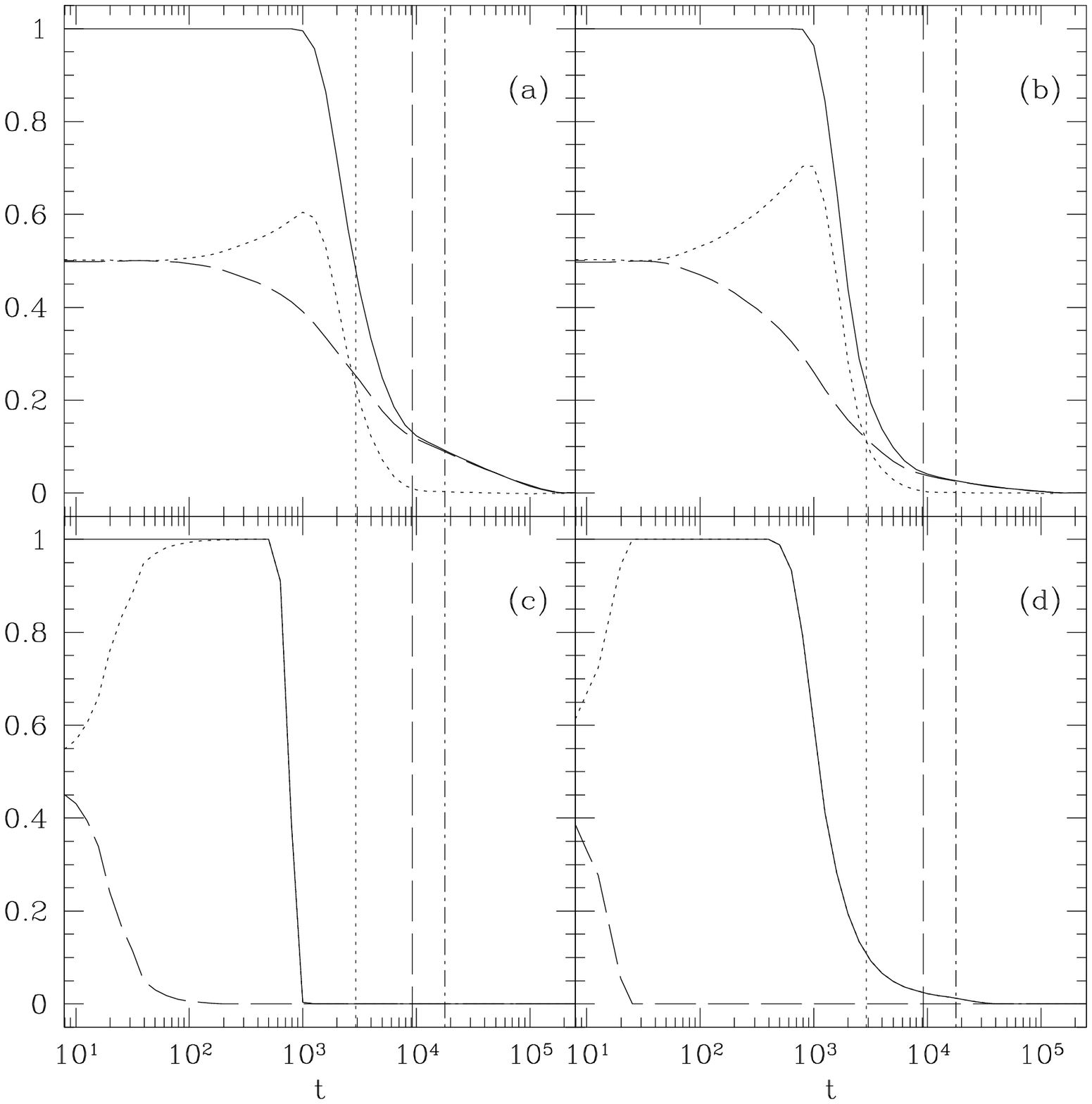}}
\caption{The fraction (to the total) of orbits not yet frictionally decayed at
time t, for a single-mass GCS population ($M_{GC}=5\times10^{-5} M$).
Solid lines refer to the whole GCS;
dashed lines to the sub-sample of clusters on
loop orbits; dotted lines to that of box orbits.
$Panel$ $(a)$: the GCS energy distribution is gaussian with $<E_0>/E_{min}=0.1$;
$Panel$ $(b)$: as in panel $(a)$, with $<E_0>/E_{min}=0.2$;
$Panel$ $(c)$: as in panel $(a)$, with $<E_0>/E_{min}=0.8$;
$Panel$ $(d)$: the energy distribution is uniform over the whole allowed range of $E_0$.
All the plots refer to the model 04.}
\label{boxloop}
\end{figure}

\begin{figure*}
\resizebox{\hsize}{!}
{\includegraphics[angle=0]{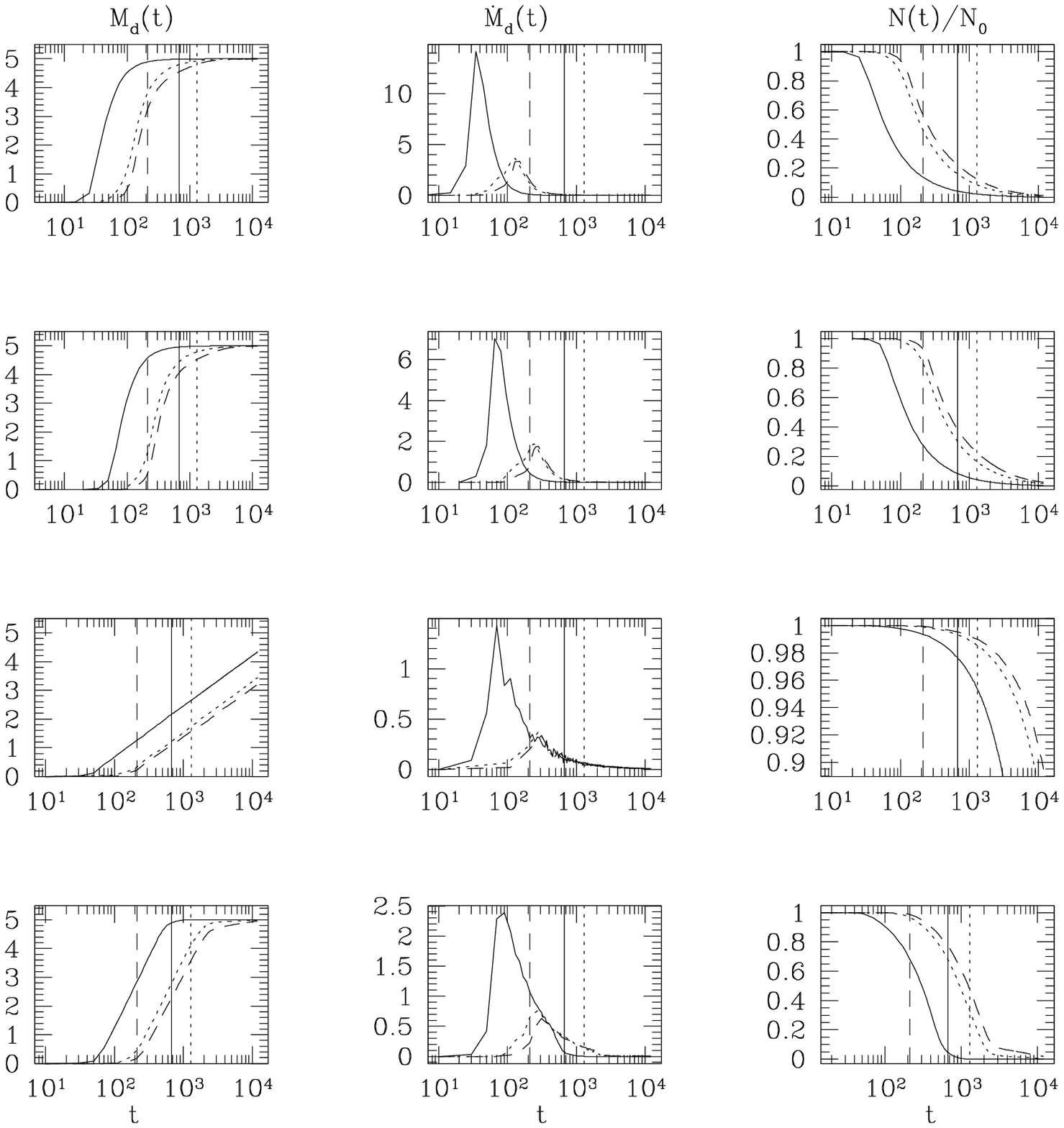}}
\caption{This figure refers to the model 21 (the model where df is more efficient) and
to three different power law mass spectra of the GCSs, whose total mass
is set to $0.05M$.
The columns refer to the time evolution of: (i) the mass ($M_d$, in units of 0.01$M$) in form of GCs
dragged into the innermost ($r\leq 0.01a$) galactic zone ($left$ $column$); (ii)
the rate of GC mass falling onto the galactic centre (in the plot $\dot{M}_d$ is in units
of $10^{-4}G^{1/2}M^{3/2} a^{-3/2}$; a rate of $1$ M$_{\sun}/yr$ corresponds to 0.15 in the scale
of the figure when $M=10^{11}$ M$_{\sun}$ and $a=1$ kpc ($central$ $column$);
(iii) the ratio (to the initial total number) of the surviving clusters ($N/N_0$)
($right$ $column$).
Solid, dotted and dashed lines refer to GCS gaussian energy distributions with
$<E>/E_{min}=0.7$, $0.2$ and $0.1$, respectively.
Vertical lines mark 1 Gyr with symbols as in Figs. \ref{plan1}.
The mass function is changed in every row. From up to bottom:
$first$ $row$: $s=0$, $M_{min}=10^{-6}$ $M$, $M_{max}=10^{-3}$ $M$;
$second$ $row$: $s=0$, $M_{min}=10^{-6}$ $M$, $M_{max}=5\times 10^{-4}$ $M$;
$third$ $row$: $s=2$, $M_{min}=10^{-6}$ $M$, $M_{max}=5\times 10^{-4}$ $M$;
$fourth$ $row$: $s=2$, $M_{min}=5\times 10^{-5}$ $M$, $M_{max}=5\times 10^{-4}$ $M$.
}
\label{Mdvst}
\end{figure*}

It is well known that GCSs in galaxies evolve in time
(Fall \& Rees 1977; Capuzzo-Dolcetta \& Tesseri 1997;
Murali \& Weinberg 1997a,b; Baumgardt 1998;
Capuzzo--Dolcetta \& Tesseri 1999; Vesperini 2001)
because of the individual internal evolution of globulars and because they, as
satellites of a galaxy, have an orbital evolution caused by both a direct and indirect
interaction with the external field, via dynamical friction
and tidal effects. Of course, in a disk galaxy the role of
the flattened gas-star distribution may be relevant and
may overcome the role of the bulge; analogously, the tidal
effect of a massive central object, if present, may overwhelm the tidal
disruptive effect of the regular large scale star distribution.

Here we limit ourselves to isolate and evaluate the role of the dynamical friction
caused by the regular triaxial distribution of stars in a
galaxy, modeled as described in Sect. \ref{Model}, on the overall GCS evolution.
This study is a first generalization of previous results (Capuzzo-Dolcetta 1993)
that showed how crucial is dynamical braking in the determination of
the observed characteristics of GCS in galaxies.
Note that neglecting evolutionary effects other than df does not introduce
an excessively large approximation relatively to the evolution of high-mass
GCS, because for them the df effect is, in any case, overwhelming (see for instance
Capriotti \& Hawley 1996).

\noindent The results we present in this section are for clusters
on planar orbits: this allows an easier presentation and
discussion of some general aspects of the GCS evolution,
without claiming to be a complete and exhaustive approach
(that will be presented in a forthcoming paper).

We assume an initial distribution of the orbital energies
of the GCs and a GCS mass spectrum, with the aim to
follow the evolution of the GCS over a time of the order of the Hubble time.

More specifically, we assume an initial population of the
different types of orbits (loops and boxes of different elongation)
given by the distribution function:
\begin{equation}
dN_{_{0,i}}=g_{_0}(M)f_{_0,i}(E) dE dM,
\end{equation}
where $i$ identifies the orbital type and the normalization
is such to have a given total mass of the GCS, indicated by $M_{GCS}$.
The total initial number of GCs is indicated by $N_0$.
For the sake of simplicity, we assume that the number
of clusters initially moving on box and loop orbits is the same, equal to $N_0/2$.
For the initial mass spectrum of the GCS, $g_{_0}(M)$, we make different choices:
$i)$ a single mass population (all the GCs have the same mass $M_{GC}=5 \times 10^{-5} M$);
$ii)$ power laws, $g_0(M)\propto M^{-s}$, choosing $s=0$ (i.e. flat spectrum) and $s=2$,
with the choices of two different low-mass cutoffs ($M_{min}= 10^{-6}M$ and
$M_{min}=5 \times 10^{-5}M$) and two different high-mass cutoffs
($M_{max}=5 \times 10^{-4} M$ and $M_{max}=10^{-3} M$).
In all cases, normalization in such to give a total mass of the GCS equal
to $5\%$ of the galaxy mass.
Of course, the choices of a single mass population and of an uniform ($s=0$) mass distribution
do not pretend to be astrophysically realistic. They were considered (together with the
more astrophysically acceptable decreasing power laws) because they allow a simplification
in the interpretation of the results, so to draw clearer and general conclusions about
the role of df as function of the satellite mass. With regard to the energy distribution, what really matters on the overall evolution of the GCS
is the average orbital energy. Actually, the lower the orbital energy the faster the orbital decay,
even if with some differences among the orbital types.
So, the details of the shape of the $f_{0,i}(E)$ are not crucial, while crucial is its average
value and the dispersion around the average. Moreover, the orbits of GCs in
a galaxy are not necessarily consistent with the galactic orbital structure (i.e. the GC velocity distribution
is not, generally, the same of the stars generating the overall potential). If we consider
that an observational determination of the GC velocity distribution is out
of the present capabilities, we conclude that there is a high degree of freedom in
the choice of the GC energy distribution.
\par\noindent In the light of the previous considerations, for
$f_{0,i}(E)$ we did the choices of a (i) flat distribution and of (ii) different gaussians.
This way, we have (i) the simplest possible $f_{0,i}(E)$, allowing an easier interpretation of the
results, and (ii) we can test the role of varying the mean orbital energy (and the dispersion
around it) of the GCS.

The fraction of ``surviving" clusters (i.e. not yet frictionally decayed) 
at various ages is shown in Fig. \ref{boxloop}, in the case of single-mass population, where the fraction of GCs moving on box
and loop orbits is given, too. In this case of single mass GCS,
the df ``disruptive" mechanism acts quite abruptly: the steep descent
of the number of surviving clusters reflects the steepness shown by
the orbital decay time vs energy relations of Figs. \ref{plan1}-\ref{plan2}.
We note the interesting feature of an inversion of the relative abundance of
GCs on box and loop orbits at sufficiently old ages in the case of initial
distributions of GC orbital energies peaked at high energies (panels $(a)$ and $(b)$
of Fig. \ref{boxloop}). The explanation of this is the action of df.
Actually, the less energetic of the sampled loop orbits become boxes
(see discussion of Fig. \ref{tevol}) inducing the peaks visible in panels $(a)$ and $(b)$;
at later ages, the relative (loop/box) abundance is, instead, influenced by that the
residual loop orbits have decay times much longer than boxes.
The above mentioned inversion is not seen at lower mean orbital energies
(panels $(c)$ and $(d)$) because df acts to transform almost all the loop orbits
into boxes before the df decay time of the initial population of box orbits.
This feature could
constitute a way to deduce information about the initial distribution
function of the GCS, whenever reliable velocity and position
data of extragalactic globular clusters will be available.

Fig. \ref{Mdvst} gives information about the mass dragged in
the innermost galactic region in form of frictionally decayed GCs ($M_d$).
The left column shows the time evolution of $M_d$, the central column gives the time-rate of 
this phenomenon ($\dot{M_{d}}$) and the right column the corresponding evolution of the
GCS population ($N/N_0$). This latter quantity (number of ``surviving" clusters to the total)
is a crucial parameter to exclude unrealistic models as those leaving too many or too few
GCs after an Hubble time.
\par\noindent We stress that $\dot{M_{d}}$ is just the quantity of mass per unit time being confined
in the innermost galactic region ($r\leq 0.01a$) by mean of decayed GCs, thus it is not pretended
to be the actual time-rate of mass accretion onto a (possible) centrally located black hole. 
Clearly, a large $\dot{M_{d}}$ is necessary to have a large accretion rate onto the compact object;
if it suffices depends on the actual accretion mechanism, that is mediated by many processes 
that deserve a careful investigation. The most likely
scenario is the one sketched by Capuzzo-Dolcetta (2002, 2003), where a dense supercluster
is formed by the merging of dynamically decayed GCs around the galactic center where a black hole
seed (even of stellar origin) should accrete spherically when the density of the supercluster 
reaches sufficiently high values. As actually shown by Capuzzo-Dolcetta (2003) on the basis of a model of
supercluster formation and evolution, both the mass rates and the time scales are totally compatible with
those required to explain the activity of galactic nuclei. In any case, much work has to be done before 
drawing firm conclusions about the role of GCS evolution in the self-regulated massive black
formation and subsequent AGN activity (Capuzzo-Dolcetta 2004).
\par\noindent The earlier steepening of the $M_d(t)$ function in the upper panel with respect
to the panel below (which obviously corresponds to an earlier peak of the mass
accretion rate) is simply due to the larger value of the average GC mass,
as implied by the greater high mass cutoff ($M_{max}$).
We note that $M_{max}$ a factor two larger corresponds to about
the same factor of reduction and increase of the time and the height of the peak
of the mass accretion rate, respectively;
actually, the direct proportionality between $\dot{M_{d}}$ and
$M_{max}$ simply reflects the behaviour of the df decay time vs $M_{GC}$.
The smaller slope of $M_d(t)$, and thus the broader mass
accretion rate in the third row (in comparison to s=0) is explained by
the $-2$ slope of the power law GCS mass function that implies a larger relative number of
light clusters, that decay later. To check it, we considered also a case with a larger $M_{min}$
($M_{min}= 5\times 10^{-5}$; bottom row).
The exclusion of light clusters (having long decay times) implies, indeed,
reaching a plateau in the centrally dragged mass (solid line in the left lowest panel)
after a growth with a steeper slope than in the panel above.

Note that, in presence of a positive correlation
between GC mass and density, lighter clusters are weaker to both tidal perturbations and
internal relaxation so to get lost in a time likely shorter than their df decay time.
Consequently, their contribution to the central mass accretion is probably overestimated
in Fig. \ref{Mdvst}.
\noindent We emphasize that Fig. \ref{Mdvst} gives important hints on the relative role played by the various
free parameters into the accretion of the galactic central region due to decayed GCs.
\noindent It results that:
$i)$ the time of the accretion burst is mainly determined by the
value of the mass of the heaviest clusters and by the mean value of their orbital energy;
$ii)$ the height of the peak of the mass accretion rate increases almost
linearly with the average mass of the GC sample and decreases with its orbital average energy (see
the various curves in each panel);
$iii)$ the length in time of the (effective)
accretion process depends mainly on the spread in the mass distribution of the GCS.

\noindent To resume, the efficiency of the accretion mechanism is maximum
for populous GCSs of low orbital energies and large individual mass.
As a consequence, few parameters are effectively relevant in the behaviour
of the decay rate in the galactic center, whose main characteristics are
the time and height of the maximum and the duration of the whole process.

\section{Conclusions}
\label{fine}
The evolution of the GCS of a galaxy is determined by
various concurrent causes. In elliptical galaxies, due to the absence of
the flattened disk component, it is easier to isolate the relevant
evolutionary effects.
A part from internal mechanisms (mass loss due to stellar evolution,
dynamical evaporation, etc.), the evolutionary effects are mainly
consisting in the collective role played by the galactic
stellar distribution through tidal distortion
and dynamical friction, as well as by the tidal interaction
between clusters and massive objects (mainly super-massive
black holes sited in the galactic centers).
In this paper we have enlarged and deepeened previous theoretical and
numerical investigations of the role of dynamical friction, extending them
to a significant range of axial ratios of triaxial
self-consistent models of core elliptical galaxies.
As galactic models, we have used 5 self consistent triaxial configurations,
computed by Statler (1987 and private communication) with
the method of the maximum entropy applied to St\"{a}ckel potentials.
The galactic models are characterized by the perfect
ellipsoid density profiles, having a central core and
decaying as $r^{-4}$ at large radii. While these profiles
do not fit with the cuspy density behaviour
observed at high resolution in many elliptical galaxies
 -so that they are not very realistic in
their central regions- they have the advantage of being
appreciably simple and apt to a deep study of
the evolution of regular orbits in triaxial potentials.
We computed a remarkably large set of orbits in the $x-y$
coordinate plane of these galaxies; fully 3D orbits have been examined, too,
but not yet deeply enough to draw fully general conclusions.

\noindent The main results of this paper may be resumed as follows:
1) df is confirmed to be important in inducing orbital
evolution of sufficiently massive globular clusters in all of the models studied;
2) triaxiality plays a role, in that {\it box} orbits (absent in the
axysimmetric case) suffer more of the frictional deceleration than the usual
{\it loop} orbits (the ratio of the loop- to box-decay time ranges from 1 to
10 in the interval of orbital energies studied,
rapidly increasing with the energy value);
in any case the initial orbital energy is the most relevant parameter, more
than the large scale galactic phase-space structure;
3) the df braking efficiency scales with the square root of
the average mass density of the galaxy (so to expect a larger variation in time of the
GCS distribution in more compact galaxies);
4) with regard to the axis ratio, we saw that for a {\it fixed average galactic
mass density}, orbits of the same type suffer more of df in more roundish galaxies.

We have studied the df effect (alone) on the global
evolution of a GCS composed by a set of globulars of different masses
and different initial conditions on position and velocity.
A relevant result is that, under general assumptions, the phase-space GCS
distribution function is heavily modified by df. Indeed, in GCSs having an
initial distribution peaked at low values of orbital energy, all the loop orbits are trasformed
into boxes before they decay completely, leaving, eventually, the galaxy
spoiled of globulars. This effect corresponds to a time evolution
of the relative abundance of loop and box orbits; unfortunately, this is
far from being observable with the present day instrumentation.

Another important consequence of dynamical friction on a GCS is the
feedback it has on the galaxy due to the accumulation of mass in form of orbitally
decayed clusters in the central regions, as it was first suggested by
Capuzzo-Dolcetta (1993) on the basis of calculations based on a single galactic model.
The present paper indicates that the results of Capuzzo-Dolcetta (1993) can be
generalized to a large interval of types of the host galaxies (of different
masses, surface brightness, axial ratios, etc.):
df has actually been able to carry into the
inner few parsecs of the galaxy an amount of massive GCs such to guarantee
the formation of a ``reservoire" of mass around a compact nucleus
in a time scale compatible with the red-shift distribution of AGNs.
Preliminary investigations of the modes and details of the actual
feeding of the central black hole by means
of this accreted mass have not been studied here, as well as the modes of conversion into
electromagnetic energy of the gravitational energy are given by Capuzzo-Dolcetta (2002, 2003).
Here we have just identified and examined the role of some of the free parameters
of the model in determining the time of the mass accretion burst, the peak of the mass accretion rate
and the length in time of the mass accretion process.
Qualitatively: the larger the typical value of the GC mass and/or the ``colder" the GCS, the
sooner the occurrence of a substantial mass accretion into the inner galaxy region;
the more numerous the GCS and/or the colder the GCS, the higher the value
of the GCS mass loss time-rate; the larger the spread of the
mass distribution of the GCS, the longer the (possible)
induced duration of the nuclear activity.

Quantitatively, we get as relevant result that -in the range of the
parameters of the orbital energy and mass
distributions of the GCSs explored here- {\it does}
exist a huge region of reasonable values corresponding to high rate of mass accretion
($\dot{M_d} >> 1$ M$_{\sun}/yr)$
and to a huge amount of mass accumulated ($M_d \geq 1/5$ $M_{GCS}$) in few $10^8$
$yr$ into the nuclear zone.
This may account for the fueling and growth of a central
massive black hole. All this convinces us of the importance to examine in a
deeper detail the actual fate of clusters interacting among themselves and
with the external field in the inner galactic regions where they are
eventually confined by dynamical friction.

\section*{Acknowledgments}
We warmly thank T. Statler for having computed for us a detailed grid of his
self-consistent triaxial models and having provided them to us in digital form.

\appendix

\section[]{Fitting formulas}


For a pratical use it is convenient to have fitting formulas for
the estimate of the df decay time, even if limiting to planar orbits.
``Compact" (even if not very accurate) expressions to
estimate the df decay time of different families of orbits
($x$-radial, less elongated box and closed loop orbits respectively)
involving the initial orbital energy ($\tilde{E_0}=E_0/E_{min}$)
and axial ratios ($b/a$ and $c/a$) are:

\begin{eqnarray}
T_{df}^{rad}=1.57 \times 10^4 \cdot \frac {\tilde{E_0}^{-0.475}}{e^{0.080 \frac {(b/a)^{-0.32}} {(c/a)}}} \times \nonumber \\
\times \bigg{(}\frac{M}{10^{11}{\rm M_{\sun}}}\bigg{)}^{- \frac {1}{2}} \bigg{(}\frac{a}{10^3 \; pc}\bigg{)}^{\frac {3} {2}} \bigg{(}\frac{M}{M_{GC}}\bigg{)} \bigg{(}{10 \over {ln \Lambda}} \bigg{)} yr
\end{eqnarray}
\begin{eqnarray}
T_{df}^{box}=1.40 \times 10^4 \cdot \frac {\tilde{E_0}^{-0.705}}{e^{0.103 \frac {(b/a)^{0.25}} {(c/a)}}} \times \nonumber \\
\times \bigg{(}\frac{M}{10^{11}{\rm M_{\sun}}}\bigg{)}^{- \frac {1}{2}} \bigg{(}\frac{a}{10^3 \; pc}\bigg{)}^{\frac {3} {2}} \bigg{(}\frac{M}{M_{GC}}\bigg{)} \bigg{(}{10 \over {ln \Lambda}} \bigg{)} yr
\end{eqnarray}
\begin{eqnarray}
T_{df}^{loop}=1.35 \times 10^3 \cdot \frac {\tilde{E_0}^{-2.134}}{e^{0.130 \frac {(b/a)^{0.36}} {(c/a)}}} \times \nonumber \\
\times \bigg{(}\frac{M}{10^{11}{\rm M_{\sun}}}\bigg{)}^{- \frac {1}{2}} \bigg{(}\frac{a}{10^3 \; pc}\bigg{)}^{\frac {3} {2}} \bigg{(}\frac{M}{M_{GC}}\bigg{)} \bigg{(}{10 \over {ln \Lambda}} \bigg{)} yr
\end{eqnarray}

More accurate fits may be obtained exploiting the existence of the
integrals of motion for the St\"{a}ckel potentials ($E_0$, $I_2$ and $I_3$, de Zeeuw 1985).
In the case of planar orbits $I_3\equiv0$, thus we have
a fitting formula depending just on $\tilde{E_0}$ and $I_2$:

\begin{equation}
\label{approx1}
T_{df}(E_0,I_2)=\bigg{[}m(\tilde{E_0})\cdot I_2+q(\tilde{E_0})\bigg{]}\bigg{(}\frac{M}{10^{11}{\rm M_{\sun}}}\bigg{)}^{- \frac {1}{2}} \bigg{(}\frac{a}{10^3 \; pc}\bigg{)}^{\frac {3} {2}} \bigg{(}\frac{M}{M_{GC}}\bigg{)} \bigg{(}{10 \over {ln \Lambda}} \bigg{)} yr
\end{equation}

\noindent
where $m$, and $q$ depends on $\tilde{E_0}$ as power laws:

\begin{equation}
\label{meq1}
m(E_0)=A \tilde{E_0}^{\alpha}
\; \; \; \; \;
q(E_0)=B \tilde{E_0}^{\beta}
\end{equation}

\noindent
with $A$, $B$, $\alpha$ and $\beta$ constants, reported in Table \ref{coeffit}.
Of course the fits are valid just in the energy range explored in our simulations:
they fail just for the lowest permitted values of $I_2$
at the lowest energies (i.e. for very elongated box and
$x$-radial orbits for energies $\tilde{E_0}=0.05$).
Under the choice of initials conditions done in this paper
($x_0=z_0=0$, $y_0\geq0$, $v_{x0}\geq0$, $v_{y0}=v_{z0}=0$)
the expression for $I_2$ (equation 49 in de Zeeuw 1985)
reduces to:

\begin{equation}
\label{i2red}
I_2=\frac{1}{2}[y_0^2-1+(b/a)^2] [2(\tilde{E_0}E_{min}-V(0,y_0,0))]^{1/2}
\end{equation}

\noindent For their use in Eq. \ref{approx1} $y_0$,
$E_{min}$ and $V(0,y_0,0)$ must enter in the units used throughout this paper.

\begin{table*}
\begin{center}
\caption{Values of the coefficients of the approximation formulae (\ref{approx1}, \ref{meq1})
for the different galactic models.}
\label{coeffit}
\begin{tabular}{|l||c|c|c|c||c|c|c|c|} \hline \hline
        & \multicolumn{4}{c|}{$I_2$}       \\ \hline
Mod     &A     &$\alpha$ &B       &$\beta$ \\ \hline
	&[$G^{-1}M^{-1}a$] & &$[G^{-1/2}M^{-1/2}a^{3/2}$] &\\ \hline
04      &29606.3  &-1.45   &90592   &-0.858  \\ \hline
06      &11763.5  &-1.60   &88878   &-0.763  \\ \hline
16      & 2749.0  &-2.11   &19377.4 &-1.051  \\ \hline
19      & 3397.2  &-1.60   &26432.6 &-0.787  \\ \hline
21      & 1981.7  &-1.60   &30142.7 &-0.656  \\ \hline \hline
\end{tabular}
\end{center}
\end{table*}

\label{lastpage}


\begin{thebibliography}{99}
\bibitem{BS00} Bak J., Statler T.S., 2000, AJ, 120, 110
\bibitem{Bau98} Baumgardt H., 1998, A\&A, 330, 480
\bibitem{Bin77} Binney J., 1977, MNRAS, 181, 735
\bibitem{BGF} Byun Y.-I., Grillmair C.J., Faber S.M., Ajhar E.A., Dressler A., Kormendy J., Lauer T.R., Richstone D., Tremaine S., 1996, AJ, 111, 1889
\bibitem{CH96} Capriotti E.R., Hawley S.H., 1996, ApJ, 464, 765
\bibitem{CD93} Capuzzo-Dolcetta R., 1993, ApJ, 415, 616
\bibitem{CT97} Capuzzo-Dolcetta R., Tesseri A., 1997, MNRAS, 292, 808
\bibitem{CT99} Capuzzo-Dolcetta R., Tesseri A., 1999, MNRAS, 308, 961
\bibitem{CV97} Capuzzo-Dolcetta R., Vignola L., 1997, A\&A, 327, 130
\bibitem{CD01} Capuzzo-Dolcetta R., 2001, in J.H. Knapen, J.E. Beckman, I.Shlosman, T.J. Mahoney, eds, ASP Conf. Ser. vol. 249, The central kpc starbursts and AGN: the La Palma connection, Astron. Soc. Pac., San Francisco, p. 237
\bibitem{CD02} Capuzzo-Dolcetta R., 2002, in E.K. Grebel and W. Brandner, eds, ASP Conf. Ser. vol. 285, Modes of Star Formation and the Origin of Field Populations, Astron. Soc. Pac., San Francisco, p. 389
\bibitem{CD03} Capuzzo-Dolcetta R., 2003, in by E. Giallongo, G. De Zotti, N. Menci, eds, Ap\&SS, Baryons in cosmic structures, Kluwer, Dordrecht in press, (astro-ph/0401541)
\bibitem{CD04} Capuzzo-Dolcetta R., 2004, in preparation
\bibitem{Ch43} Chandrasekhar S., 1943, ApJ, 97, 255
\bibitem{CVM} Cora S.A., Vergne M.M., Muzzio J.C., 2001, ApJ, 546, 165
\bibitem{dZ85} de Zeeuw P.T., 1985, MNRAS, 216, 273
\bibitem{dZL85} de Zeeuw P.T., Lynden-Bell D., 1985, MNRAS, 215, 713
\bibitem{dZM83} de Zeeuw P.T., Merritt D., 1983, ApJ, 267, 571
\bibitem{Deh93} Dehnen W., 1993, MNRAS, 265, 250
\bibitem{FTA96} Faber S.M., Tremaine S., Ajhar E.A., Byun Y.-I., Dressler A., Gebhardt K., Grillmair C., Kormendy J., Lauer T.R., Richstone D., 1997, AJ, 114, 1771
\bibitem{FR77} Fall S.M., Rees M.J., 1977, MNRAS, 181, 37
\bibitem{FIdZ} Franx M., Illingworth G., de Zeeuw P.T., 1991, ApJ, 383, 112
\bibitem{GRAL} Gebhardt K., Richstone D., Ajhar E.A., Lauer T.R., Byun Y.-I., Kormendy J., Dressler A., Faber S.M., Grillmair C., Tremaine S., 1996, AJ, 112, 105
\bibitem{GB85} Gerhard O., Binney J., 1985, MNRAS, 216, 467
\bibitem{Her90} Hernquist L., 1990, ApJ, 356, 359
\bibitem{HB01} Holley-Bockelmann K., Mihos J.C., Sigurdsson S., Hernquist L., 2001, ApJ, 549, 862
\bibitem{Lan60} Lance G.N., 1960, Numerical methods for high-speed computers, Iliffe \& Sons, London, p. 56
\bibitem{LABD} Lauer T.R., Ajhar E.A., Byun Y.-I., Dressler A., Faber S.M., Grillmair C., Kormendy J., Richstone D., Tremaine S., 1995, AJ, 110, 2622
\bibitem{L56} Lindblad, B., 1956, Stockholm Obs. Ann., 19, 2
\bibitem{MF96} Merritt D., Fridman T., 1996, ApJ, 460, 136
\bibitem{MV96} Merritt D., Valluri M., 1996, ApJ, 471, 82
\bibitem{MW97a} Murali C., Weinberg M.D., 1997a, MNRAS, 288, 767
\bibitem{MW97b} Murali C., Weinberg M.D., 1997b, MNRAS, 291, 717
\bibitem{OBS} Ostriker J.P., Binney J., Saha P., 1989, MNRAS, 241, 849
\bibitem{Ost} Ostriker J.P., 1988, in J.E. Grindlay, A.G. Davis Philip, eds, IAU Symp. n.126, The Harlow Shapley Symposium on Globular Cluster Systems in Galaxies, Reidel, Dordrecht, p.271
\bibitem{PCV} Pesce E., Capuzzo-Dolcetta R., Vietri M., 1992, MNRAS, 254, 466
\bibitem{Ryd} Ryden B.S., 1996, ApJ, 461, 146
\bibitem{Sper} Spergel D.N. et al, 2003 ApJS, 148, 175
\bibitem{S87} Statler T.S., 1987, ApJ, 321, 113
\bibitem{S91} Statler T.S., 1991, AJ, 102, 882
\bibitem{Sch79} Schwarzschild M., 1979, ApJ, 232, 236
\bibitem{Sch82} Schwarzschild M., 1982, ApJ, 263, 599
\bibitem{Schwei79} Schweizer F., 1979, ApJ, 233, 23
\bibitem{WMAP} Spergel D.N. et al., 2003, ApJS, 148, 175
\bibitem{SMFM} Stiavelli M., Miller B.W., Ferguson H.C., Mack J., Whitmore B.C., Lotz J.M., 2001, AJ, 121, 1385
\bibitem{TOS} Tremaine S.D., Ostriker J.P., Spitzer L.Jr., 1975, ApJ, 196, 407
\bibitem{TM95} Tremblay B., Merritt D., 1995, AJ, 110, 1039
\bibitem{V01} Vesperini E., 2001, MNRAS, 332, 247
\bibitem{W24} Weinacht J., 1924, Math. Ann., 91, 279


\end{thebibliography}
\end{document}